\documentclass[12pt,a4paper,dvips]{article}            

\usepackage{a4wide}
\usepackage[english]{babel}
\usepackage[dvips]{graphics}

\newcommand{\be}{\begin{equation}}
\newcommand{\ee}{\end{equation}}
\newcommand{\beqn}{\begin{eqnarray}}
\newcommand{\eeqn}{\end{eqnarray}}
\newcommand{\bes}{\begin{eqnarray*}}
\newcommand{\ees}{\end{eqnarray*}}
\newcommand{\beqns}{\begin{eqnarray*}}
\newcommand{\eeqns}{\end{eqnarray*}}

\newcommand{\rmd}{\mbox{d}}
\newcommand{\rme}{\mbox{e}}
\newcommand{\rmi}{\mbox{i}}
\newcommand{\cE}{{\cal E}}

\newcommand{\srmi}{\mbox{\scriptsize i}}

\newcommand{\dd}[2]{{\rmd{#1}\over\rmd{#2}}}
\newcommand{\pdd}[2]{{\partial{#1}\over\partial{#2}}}

\begin{document}

\gdef\journal#1, #2, #3, #4#5#6#7{{#1~}{\bf #2}, #3 (#4#5#6#7)}
\gdef\ibid#1, #2, #3#4#5#6{{\bf #1} (#3#4#5#6) #2}
\begin{center}
{\Large\bf Quantum Mechanics of a Particle
with Two Magnetic Impurities}\\[1.5cm]
{\large Stefan Mashkevich}\footnote{mash@mashke.org}\\[0.1cm]
Physics Department,
Taras Shevchenko Kiev National University,
03022 Kiev, Ukraine\\[0.4cm]
{\large Jan Myrheim}\footnote{jan.myrheim@phys.ntnu.no}\\[0.1cm]
Department of Physics,
The Norwegian University of Science and Technology,\\
N--7034 Trondheim, Norway\\[0.4cm]
{\large St\'ephane Ouvry}\footnote{ouvry@ipno.in2p3.fr}\\[0.1cm]
Laboratoire de Physique Th\'eorique et Mod\`eles
Statistiques\footnote{Unit\'e de
Recherche de l'Universit\'e Paris-Sud associ\'ee au CNRS, UMR 8626}\\
B\^at. 100, Universit\'e Paris-Sud, 91405 Orsay, France
\end{center}

\vskip 0.5cm
\centerline{\large \bf Abstract}
\vskip 0.2cm

A two-dimensional quantum mechanical system consisting of a particle
coupled to two magnetic impurities of different strengths,
in a harmonic potential, is considered.
Topological boundary conditions at impurity locations
imply that the wave functions are linear combinations
of two-dimensional harmonics.
A number of low-lying states are computed numerically,
and the qualitative features of the spectrum are analyzed.

\vskip 1cm
\noindent

PACS numbers:
03.65.-w, 05.30.-d, 11.10.-z, 05.70.Ce

Keywords:
Magnetic Impurities, Anyons, Quantum Hall effect, Winding

\newpage
\section{Introduction}

A distinctive feature of the topology of two-dimensional space
is that a pointlike object renders it multiply connected.
A closed path taking a particle around the object
is characterized by a winding number, showing how many times
the object has been encircled;
and in the path-integral formalism, 
paths with different winding numbers
may carry different overall phase factors.

A simple physical incarnation
of this general principle is a system of charged particles
in the presence of ``magnetic impurities'', i.e.~magnetic
flux points (two-dimensional projections of flux tubes)
at given points in space. There is no classical interaction force,
since the magnetic field outside an impurity vanishes; however,
the electromagnetic vector potential being
topologically nontrivial, the
dynamics of the particle is influenced (the Aharonov-Bohm effect \cite{AB}).
Identifying impurities with particles themselves leads to
the concept of   anyon
statistics \cite{LM}, whereby the wave function
acquires a nontrivial phase factor when the positions
of two particles are continuously interchanged.
The anyon model is not exactly solvable
for more than two particles,
because the usual representation of the multiparticle wave function
(which is now a multivalued function of its complex arguments)
in terms of products of single-particle functions is not possible.

Obviously, the two-dimensional model with magnetic impurities may be viewed
as a projection of a three-dimensional system in which
the impurities are infinite solenoids,
and the longitudinal motion is free.

Topological models of this kind are interesting
because of their relevance to
the fractional quantum Hall effect \cite{Hall}, which
is a two-dimensional phenomenon and where the elementary
excitations are believed to be anyonic.
For a particle coupled to a random Poissonian distribution 
of magnetic impurities, calculations of the
density of states averaged over the positions
of the impurities have been performed \cite{density},
with some interesting qualitative conclusions.
The model is no less fascinating, however, as
an example of quite a generic quantum-mechanical system
in multiply connected space.

Last but not least, there is a direct connection with the problem
of winding number distribution of random paths \cite{winding}.
In fact, the partition function of one particle in the presence
of $N$ impurities is essentially a Fourier transform of
the sequence $P_{m_1m_2...m_N}$ of probabilities that
a random closed path winds $m_k$ times around point $k$
for all $k=1,\ldots,N$.
Since it is known from past experience \cite{3virial} that
the partition function of such systems
can be computed with precision better
by several orders of magnitude
(with comparable computational time expense)
by means of finding the spectrum directly
rather than by Monte Carlo simulating the winding number distribution,
one tends to believe that finding the levels is the right way to go
in pursuit of a good numerical, and possibly analytic,
estimate of the probability distribution in question.

The system of one particle coupled to one impurity is trivial (standard A-B).
The kinetic angular momentum with respect to the impurity acquires a
fractional part equal to the magnetic flux modulo the flux quantum, and the
energy levels, which depend on the momentum, shift accordingly. In this paper
we consider the problem of one particle coupled to two impurities \cite{Stov}.
A harmonic potential is used as the most convenient long-distance regulator.
There appears a new dimensionless parameter, the ratio of the distance between
the two impurities to the harmonic length. (In the thermodynamic limit, when
the harmonic frequency vanishes, it is the ratio of that distance to the
thermal wavelength that becomes relevant.) Searching for the wave
function as a linear combination of two-dimensional harmonics and imposing
boundary conditions at the positions of the impurities boils down to the
standard procedure: there arises a system of linear equations for the
coefficients, and the levels are those values of energy at which the
determinant vanishes.

We find numerically the ground state and the first two
excited states and observe how their energies depend
on the strengths of the impurities
(i.e.~the values of the fluxes they carry)
and the distance between them. 
At certain conditions there is a nonperturbative behavior
at vanishing distance, whose nature is the same as in the case
of the A-B ground state \cite{pert}:
a nonsingular wave function turns into a singular one.
The present calculation is a starting step in the
evaluation of the partition function, with the subsequent
inference of the two-point winding number distribution.

\section{Two-dimensional harmonics and boundary conditions}

The single-particle two-dimensional harmonic oscillator Hamiltonian
is ($\hbar = 1$)
\be
H=-\frac{1}{2m}\left(
\pdd{^2}{r^2}+{1\over r}\,\pdd{}{r}
+{1\over r^2}\,\pdd{^2}{\varphi^2}
\right)
+\frac{m\omega^2}{2}\,r^2\;.
\ee
Upon introducing the dimensionless length $\rho = \sqrt{m\omega}r$
and energy $\cE=E/\omega$ and separating the wave function
as $\psi(\rho,\varphi)=f(\rho)\,\rme^{\srmi l\varphi}$,
the energy eigenvalue equation $H\psi=E\psi$ reduces to the radial equation
\be
\left(
\dd{^2}{\rho^2}+{1\over\rho}\,\dd{}{\rho}-{l^2\over\rho^2}
-\rho^2+2\cE\right)f(\rho)=0\;.
\ee
Its two linearly independent solutions can be expressed
in terms of the confluent hypergeometric functions \cite{AS}:
\beqn
f_{l,\cE}(\rho) & = &
\rho^{|l|}\,\rme^{-{\rho^2\over 2}}\,M(a,b,\rho^2)\;, \\
g_{l,\cE}(\rho) & = &
\rho^{|l|}\,\rme^{-{\rho^2\over 2}}\,U(a,b,\rho^2)\;,
\eeqn
where
\be
a={|l|+1-{\cal E}\over 2}\;,\qquad
b=|l|+1\;.
\label{ab}
\ee
$M(a,b,\rho^2)$ is regular at $\rho=0$ but in general grows exponentially
at $\rho\to\infty$; on the contrary, $U(a,b,\rho^2)$ falls off at
infinity but is singular at the origin. It is only in the special
case of $a=-n$, where $n=0,1,2,\ldots$, that the two solutions actually
coincide and represent a wave function that is regular both at
zero and infinite $\rho$; hence, the familiar
two-dimensional harmonic spectrum
\be
{\cal E} = 2n + |l|+1\;,
\label{E}
\ee
with the integer angular momentum $l=0,\pm1,\pm2,\ldots$
(the wave function being single-valued).

When the particle encircles an impurity with flux $\Phi$ at the origin, one
can always work in the singular gauge where the flux has been erased but the
wave function acquires an extra factor of $\exp(\rmi\alpha\varphi)$,
where the ``impurity strength'' is, by definition,
$\alpha = \Phi/\Phi_0$, with the flux quantum $\Phi_0 = 2\pi/e$.
Thus, moving the particle around the origin brings up a
phase factor of $\exp(2\rmi\alpha\pi)$; for fractional $\alpha$,
the wave function becomes multivalued.
This is achieved by a shift
$l\mapsto l+\alpha$; accordingly, the spectrum becomes
\be {\cal E} = 2n + |l+\alpha|+1\;,
\label{Efrac}
\ee
Note that the
phase factor---and the spectrum---is periodic in $\alpha$ with period 1;
i.e., it is only the
fraction of the flux quantum that matters.

Let us now place two impurities on the plane:
one of strength $\alpha_1$
at the point with polar coordinates $(\rho_0, 0)$
and another one of strength $\alpha_2$ at $(\rho_0, \pi)$.
The most general particle wave function
inside the circle $\rho = \rho_0$ is
\be
\psi_-(\rho,\varphi)=\sum_{l=-\infty}^{\infty}
c_{l}\,f_{l,\cE}(\rho)\,\rme^{\srmi l\varphi}\;.
\label{psim}
\ee
Outside the same circle, adding $2\pi$ to $\varphi$ amounts to
encircling both impurities 1 and 2, which should yield
a phase factor of $\exp[2\rmi(\alpha_1+\alpha_2)\pi]$.
The wave function, therefore, must be
\be
\psi_+(\rho,\varphi)=\sum_{l=-\infty}^{\infty}
d_{l}\,g_{l+\alpha_1+\alpha_2,\cE}(\rho)\,
\rme^{\srmi(l+\alpha_1+\alpha_2)\varphi}\;.
\label{psip}
\ee
On the circle $\rho=\rho_0$ we must require continuity of the function
$\psi(\rho,\varphi)$ and its normal, i.e.~radial derivative. 
\begin{figure}
\begin{center}
\includegraphics{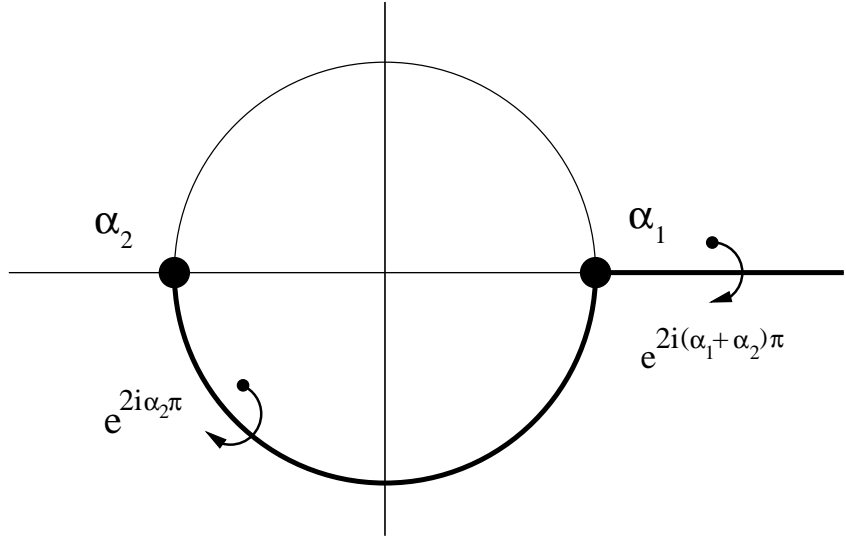}
\end{center}
\caption{The two branch cuts and associated phase differences.}
\label{cuts}
\end{figure}
There remain, however, the boundary conditions associated with
encircling the individual impurities; those cannot be satisfied
with one global relation between $\psi_-(\rho_0,\varphi)$ and 
$\psi_+(\rho_0,\varphi)$.
Rather, one has to introduce branch cuts such that encircling
zero, one, or both impurities would involve crossing
different combinations of the cuts, and impose
phase differences between the opposite
sides of each cut. Indeed, if the two cuts
are made as shown on Fig.~\ref{cuts},
the boundary conditions are satisfied.
E.g., encircling impurity 1 counterclockwise yields
a phase factor of
$\exp[-2\rmi\alpha_2\pi] \times \exp[2\rmi(\alpha_1+\alpha_2)\pi]
= \exp[2\rmi\alpha_1\pi]$.
Thus, the boundary conditions are
\beqn
\psi_-(\rho_0,\varphi) = \psi_+(\rho_0,\varphi)\qquad&&(0\le\varphi<\pi)\;; \\
\psi_-(\rho_0,\varphi) = \rme^{-2\srmi\alpha_2\pi}\psi_+(\rho_0,\varphi)
\qquad&&(\pi\le\varphi<2\pi)\;,
\eeqn
or, substituting the explicit expressions (\ref{psim})--(\ref{psip}),
\beqn
\sum_{l=-\infty}^{\infty}
c_{l}\,f_{l,\cE}(\rho_0)\,\rme^{\srmi l\varphi}
=
\sum_{l=-\infty}^{\infty}
d_{l}\,g_{l+\alpha_1+\alpha_2,\cE}(\rho_0)\,
\rme^{\srmi(l+\alpha_1+\alpha_2)\varphi}
\qquad&&(0\le\varphi<\pi)\;; 
\label{bc1}
\\
\sum_{l=-\infty}^{\infty}
c_{l}\,f_{l,\cE}(\rho_0)\,\rme^{\srmi l\varphi}
=\rme^{-2\srmi\alpha_2\pi}
\sum_{l=-\infty}^{\infty}
d_{l}\,g_{l+\alpha_1+\alpha_2,\cE}(\rho_0)\,
\rme^{\srmi(l+\alpha_1+\alpha_2)\varphi}
\qquad&&(\pi\le\varphi<2\pi)\;.
\label{bc2}
\eeqn
Exactly the same conditions have to hold for the normal derivatives
$\partial\psi_\pm(\rho,\varphi)/\partial\rho$, implying
\beqn
\sum_{l=-\infty}^{\infty}
c_{l}\,f_{l,\cE}'(\rho_0)\,\rme^{\srmi l\varphi}
=
\sum_{l=-\infty}^{\infty}
d_{l}\,g_{l+\alpha_1+\alpha_2,\cE}'(\rho_0)\,
\rme^{\srmi(l+\alpha_1+\alpha_2)\varphi}
\qquad&&(0\le\varphi<\pi)\;; \label{bc3}
\\
\sum_{l=-\infty}^{\infty}
c_{l}\,f_{l,\cE}'(\rho_0)\,\rme^{\srmi l\varphi}
=\rme^{-2\srmi\alpha_2\pi}
\sum_{l=-\infty}^{\infty}
d_{l}\,g_{l+\alpha_1+\alpha_2,\cE}'(\rho_0)\,
\rme^{\srmi(l+\alpha_1+\alpha_2)\varphi}
\qquad&&(\pi\le\varphi<2\pi)\;
\label{bc4}
\eeqn
[$f_{l,\cE}'(\rho) \equiv \partial f_{l,\cE}(\rho)/\partial\rho$].
These equations depend on energy as a parameter,
and energy levels are those
values of $\cE$ for which all four can be satisfied simultaneously.

\section{Numerical procedure}

In order to project out single Fourier components, one
should multiply Eqs.~(\ref{bc1})--(\ref{bc4})
by $\rme^{-\srmi k\varphi}$, for integer $k$,
and integrate over $\varphi$ from $0$ to $2\pi$.
In practice, however, the $l$ summation will be truncated
to a finite number of terms, say $2N$. It is impossible
to satisfy the boundary conditions at all values of $\varphi$
by adjusting the finite number of coefficients.
A better approach, leading to greater numerical
precision, is to impose those conditions
at $2N$ uniformly distributed
discrete points $\varphi_n=(2n+1)\pi/(2N)$,
where $n=0,1,2,\ldots,2N-1$ (cf.~\cite{3virial}).
Integration over $\varphi$ then gets replaced
with summation over $\varphi_n$.
Equations (\ref{bc1}) and (\ref{bc2}) turn into one,
\be
2Nc_k\,f_{l,\cE}(\rho_0)
=
\rme^{\srmi\alpha_1\pi}\sum_{l}
{\sin(\alpha_1\pi)+(-1)^{l-k}\,\sin(\alpha_2\pi)
\over\sin{(l+\alpha_1+\alpha_2-k)\pi\over 2N}}\,
d_{l}\,g_{l+\alpha_1+\alpha_2,\cE}(\rho_0)\;,
\label{int-bc}
\ee
as do (\ref{bc3}) and (\ref{bc4}).
%
%
Eliminating the $c_k$'s yields
\be
\sum_l A_{kl}(\cE)\,d_l = 0\;,
\ee
with the matrix element
\be
A_{kl}(\cE) =
{\sin(\alpha_1\pi)+(-1)^{l-k}\,\sin(\alpha_2\pi)
\over2N\sin{(l+\alpha_1+\alpha_2-k)\pi\over 2N}}
\left[
f_{l,\cE}(\rho_0)\,g_{l+\alpha_1+\alpha_2,\cE}'(\rho_0)
-f_{l,\cE}'(\rho_0)\,g_{l+\alpha_1+\alpha_2,\cE}(\rho_0)\right]\;.
\label{matel}
\ee
The $l$ summation may be over any $2N$ values (not even necessarily
sequential): any basis can be chosen for expanding
the wave function, but the more adequate the choice, the more accurate
the levels found. At least for low-lying states, the right choice
is to have the values $l+\alpha_1+\alpha_2$ group as closely
around zero as possible.

In the $N\to\infty$ limit, the denominator in Eq.~(\ref{matel})
gets replaced with $(l+\alpha_1+\alpha_2-k)\pi$, which is
precisely what one would get if Eqs.~(\ref{bc1})--(\ref{bc4})
were integrated over $\varphi$ rather than summed over $\varphi_n$.

The numerical procedure is simply to scan a desired interval
of energy, for given $\alpha_1$, $\alpha_2$, and $\rho_0$,
to find the values that nullify the determinant of $||A_{kl}||$.

There is an evident fourfold symmetry in the $(\alpha_1,\alpha_2)$ space.
Apart from the $(\alpha_1,\alpha_2) \mapsto
(\alpha_2,\alpha_1)$ invariance, there is $P$-symmetry (reversal
of signs of both $\alpha$'s) as well as periodicity
($\alpha_j \mapsto \alpha_j+1$). Taken together, these
imply invariance with respect to $(\alpha_1,\alpha_2) \mapsto
(1-\alpha_2,1-\alpha_1)$. Hence, the spectrum is symmetric
with respect to both diagonals of the square $0\le\alpha_{1,2}\le1$,
and it is enough to confine oneself to any one of the four triangles
formed by those diagonals.

A big improvement in precision can be
gained by finding one and the same level
at different values of $N$ and then extrapolating to $N\to\infty$.
In fact, comparing the results for different $N$ is the
only reliable way to estimate the error caused by the truncation.
Convergence in $N$ is better for wave functions that are less singular
at impurity positions, since it is difficult to represent  singular
functions with  finite Fourier series. The leading-term behavior of
a generic wave function near an impurity of strength $\alpha$
($0\le\alpha\le1$), with the distance from the impurity
$r\to0$, is $r^{1/2-|\alpha-1/2|}$.
Convergence, therefore, is expected to be the best
at $\alpha=1/2$ and the worst
near the endpoints $\alpha=0$ and $\alpha=1$.

This is indeed what is observed. The finite-$N$ data,
for $N$ not too small, are found to be quite well described
by an empiric formula
\be
\cE(N) = \cE + c\,N^{-\nu}\;,
\label{nfit}
\ee
with the exponent $\nu$ depending rather weakly on $\rho_0$,
but strongly on the $\alpha$'s (Table 1).

\begin{center}
\begin{tabular}{|c||l|l|l||l|l|l|}
\hline
 & \multicolumn{3}{c||}{$\rho_0=0.75$} &
\multicolumn{3}{c|}{$\rho_0=1.5$} \\ \cline{2-7} 
\raisebox{1.5ex}[0mm]{$\alpha_1, \alpha_2$} &
\multicolumn{1}{c|}{$\cE$} &
\multicolumn{1}{c|}{$\nu$} & 
\multicolumn{1}{c||}{$c$} & 
\multicolumn{1}{c|}{$\cE$} &
\multicolumn{1}{c|}{$\nu$} &
\multicolumn{1}{c|}{$c$} \\
\hline
 0.1, 0.1 &
1.147 & 0.10 & -0.153 & 
1.027 & 0.09 & -0.030 \\
 0.3, 0.2 &
1.280 & 0.41 & -0.357 &
1.043 & 0.41 & -0.077 \\
 0.5, 0.5 &
1.442 & 1.01 & -0.773 &
1.060 & 0.99 & -0.272 \\
 0.7, 0.0 &
1.135 & 0.67 & -0.039 &
1.023 & 0.82 & -0.029 \\
 0.9, 0.3 &
1.195 & 0.33 & -0.200 &
1.032 & 0.38 & -0.051 \\
\hline
\end{tabular}
\vskip 0.2cm
Table 1: The extrapolated ground state energy and the convergence
rate, from a fit to Eq.~(\ref{nfit}).
The values of $N=40$, 80, 160, 320, 640 were used for the fitting.
\end{center}

\begin{figure}
\begin{center}

\scalebox{0.75}{
\includegraphics{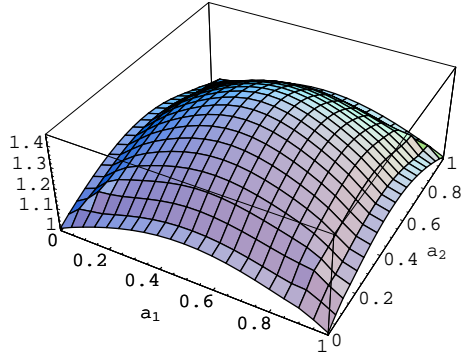}
}

(a)\\[0.3cm]
\scalebox{0.75}{
\includegraphics{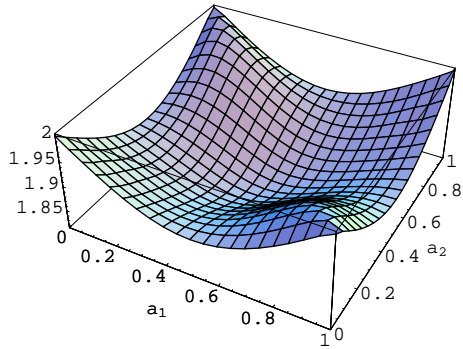}
}

(b)\\[0.3cm]
\scalebox{0.75}{
\includegraphics{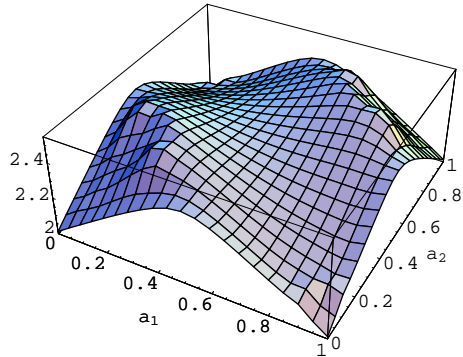}
}

(c)

\caption{The energies of the ground state (a)
and of the first two excited states (b,c)
as functions of $\alpha_1$, $\alpha_2$, at $\rho_0=0.75$.}
\label{Ea1a2}
\end{center}
\end{figure}

The same technique was used in Ref.~\cite{3virial} in
the treatment of the three-anyon problem, the whole numerical
algorithm being rather similar to the present one.
The convergence there is much better close to the fermionic
end $\alpha=1$, where the wave functions are less singular;
a certain symmetry between $\alpha$ and $1-\alpha$
enabled one to altogether escape the region $0<\alpha<1/2$
with poor convergence. The trick does not work here, however:
for all four values of $(\alpha_1, \alpha_2)$ equivalent
due to symmetry, convergence is about the same.

\section{Results and discussion}

Shown on Fig.~\ref{Ea1a2} are the energies of the three lowest
states as functions of the impurity strengths at a fixed
value of distance between impurities.
The values $N=80$, 160, 320, 640 were used for the extrapolation
by means of fitting to Eq.~(\ref{nfit}).
By qualitative analogy with the relative motion of two anyons,
introducing the impurities with positive strength
increases the (algebraic) mean value
of the particle's kinetic angular momentum.
Therefore, the energy increases if the angular momentum
for zero impurity strengths is nonnegative, and decreases
if it is negative. (Since in this case, eigenstates of
the Hamiltonian are not eigenstates of the angular momentum,
neither, in general, are the ``correct'' basis states at
$\alpha_1=\alpha_2=0$, into which the level splits as
the impurities are turned on. However, the statement holds
as pertaining to the mean value.)

\begin{figure}
\begin{center}
\includegraphics{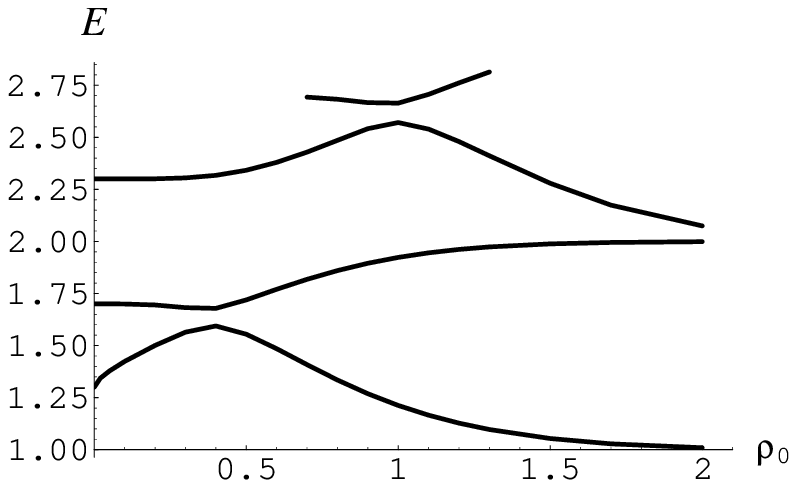}

(a)

\includegraphics{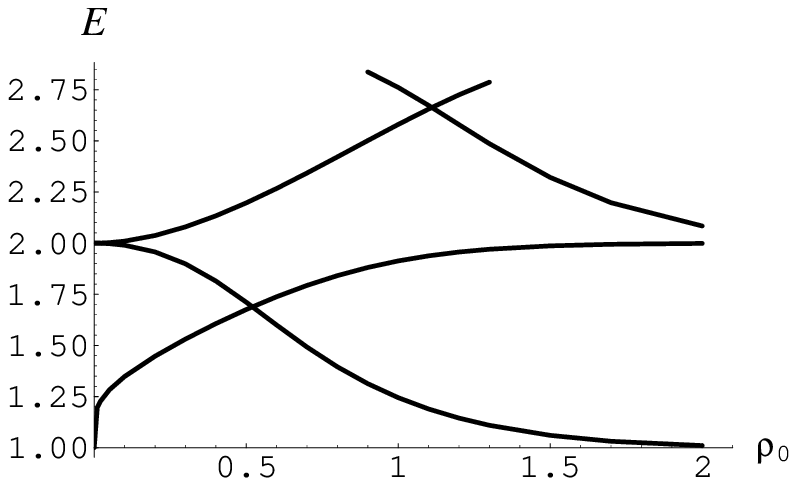}

(b)

\caption{The energies of the ground state
and of the first two excited states 
as functions of $\rho_0$:
(a) $\alpha_1=0.4$, $\alpha_2=0.3$,
(b) $\alpha_1=\alpha_2=0.5$.}
\label{Erho}
\end{center}
\end{figure}

A separate issue is the dependence of the levels on the distance
between impurities.
At $\rho_0=0$, the levels are given by Eq.~(\ref{Efrac})
with $\alpha = \alpha_1+\alpha_2$.
At infinite $\rho_0$, one has
the single-particle oscillator spectrum (\ref{E}).
For very small $\alpha_1$ and $\alpha_2$, there is
a clear mapping between the two spectra, and with the
change of $\rho_0$, each state
will slowly interpolate into its counterpart.
As the impurity strengths increase, levels come closer
to crossing each other. However,
as long as $\alpha_1\ne\alpha_2$,
there can be only avoided crossings, since there is
no symmetry in the system.
(Reflection with respect to the $x$ axis reverses the
signs of the $\alpha$'s.) A generic case
is depicted on Fig.~\ref{Erho}~(a).

True crossings become possible for
$\alpha_1=\alpha_2$, when there is $\mathrm{C}_2$ symmetry;
cf.~Fig.~\ref{Erho}~(b).
These will turn into avoided crossings when
that symmetry is broken in any way---%
say, resulting from a change, no matter how small,
in the position or strength of one of the impurities.
Continuity is still maintained, however:
If the change is small, so will be the minimal distance
between the levels, and their wave functions will
be mutually ``flipping'' in the narrow region near the avoided
crossing---so that the overall picture will be little
different from a true crossing.

There is one more peculiarity in our second example,
which arises whenever $\alpha_1+\alpha_2$ is an integer---%
in this case, 1.
(The spectrum is then the same at zero and infinite $\rho_0$.)
For small $\rho_0$, the ground state energy behaves nonperturbatively,
$\left.\partial\cE_0/\partial\rho_0\right|_{\rho_0=0}$ turning infinite.
This is due to the fact that in this and only this case,
the wave function of the ground state (and indeed of all
states with zero angular momentum) is not singular for $\rho_0=0$
but becomes singular when the two impurities are split up.
The same reason is responsible for the breakdown
of perturbation theory which has been long known
in pertinence to the A-B problem:
A direct attempt to treat a weak flux
as a perturbation and to build up an expansion in $\alpha$
fails for the ground state,
because its singular wave function, $r^\alpha$ with
a small $\alpha$, cannot be obtained from the regular unperturbed
wave function, $r^0$, by perturbation.
A special treatment is needed in order to enable
perturbation theory to produce the correct answer \cite{pert}.

States with nonzero angular momentum, on the contrary,
have $\left.\partial\cE_0/\partial\rho_0\right|_{\rho_0=0}=0$,
since their wave functions vanish at the origin fast enough.

One final remark is in order concerning the case 
of big distances.
At any $\rho_0 \gg 1$, there will exist states
whose wave functions are situated almost totally inside the
$\rho_0$ circle, as well as ones (high enough in the spectrum)
whose wave functions are almost totally outside.
Thus, while it is true that any given level will,
at big enough $\rho_0$, assume its integer value
which it has in the absence of impurities,---%
at any given $\rho_0$, no matter
how big, there exist infinitely many levels whose energies
tend to their $\rho_0=0$ values.
As is commonly the case when there
are two competing parameters, the order of limiting
transitions is crucial.
In practice, an energy cutoff will be ensured
by a finite temperature; and the presence of the impurities
will not be felt when the distance between them is much
greater than the thermal wavelength.

To conclude, we have construed a numerical algorithm
for the solution of the problem of
a particle coupled to two magnetic impurities in two dimensions,
and found several low-lying states.
It would be quite straightforward to generalize this
to any number of impurities on a circle:
The integrated boundary conditions would still have the
form akin to (\ref{int-bc}), with just the matrix elements
modified. Of course, for more than two impurities,
placing them all on one circle is a loss of generality
(not so for two impurities in the thermodynamic limit,
when the center of the harmonic potential ceases to exist).
With several impurities not on one circle,
the boundary conditions would be more complicated,
but in principle still solvable.

Evaluating the partition function of the system considered here
and inferring the probability distribution of winding numbers
of a particle around two fixed points on a plane
is left for future work.

\end{document}